\documentstyle[seceq]{ptptex}

\newcommand{\EQ}{\begin{equation}}
\newcommand{\EN}{\end{equation}}
\newcommand{\EQA}{\begin{eqnarray}}
\newcommand{\EQN}{\end{eqnarray}}
\newcommand{\EQAN}{\begin{eqnarray*}}
\newcommand{\EQNN}{\end{eqnarray*}}
\newcommand{\e}{{\rm e}}

\newcommand{\Tr}{{\rm Tr}}

\title{
Holography in the Large-$J$ Limit \\of AdS/CFT Correspondence 
and Its Applications
 }

\author{
Tamiaki {\sc Yoneya}\footnote{E-mail: tam@hep1.c.u-tokyo.ac.jp. 
This article is based on a talk given by the 
present author in the symposium, 
``Frontiers of Quantum Science", YITP, Kyoto University, February, 
2005. Though the talk was given more than a year ago, 
some of recent results are taken into account in this written 
version which is to be published in the Proceedings. }
}

\inst{
Institute of Physics, University of Tokyo, Komaba 153-8902, Tokyo
}

\abst{
 We give a brief expository discussion on the holographic 
correspondence of correlation functions 
in the large-$J$ limit of AdS/CFT conjecture.  
We first review our proposals on the interpretation of 
the so-called GKPW relation in the large-$J$ limit or BMN limit, which are 
based upon a tunneling picture in 
relating the AdS bulk to its boundary. 
Some concrete results, explicitly confirming    
our picture, are  summarized. We then proceed to comment on 
various issues related to this subject, such as extension 
of the present picture to nonconformal D$p$-brane 
backgrounds,  the 
correlators of deformed Wilson loops, 
spinning-string/spin-chain correspondence, and the inclusion of 
higher string-loop effects.  In particular, as for the deformation of 
Wilson loops, we present a typical tunneling world-sheet 
solution which can be used for direct derivation of the 
expectation values of deformed Wilson loops 
following our picture.  
}

\begin{document}

\maketitle

\section{Introduction}
The AdS/CFT (or more generally {\it string/gauge})
 correspondence has been one 
of central themes in recent developments 
of string theory, since its first proposal by Maldacena \cite{maldacena}
 in 1997. 
It is a remarkable conjecture that makes possible to 
connect the gravitational physics of strings in bulk spacetimes  
to that of non-gravitational field theories living on their boundaries.  
Establishing such a `holographic' correspondence would mean that 
we may achieve an entirely new synthesis 
of various different field theories using physics of strings and 
branes. On one hand, various old attempts, starting from the 70's,  towards the derivation of string picture from gauge theories, are precursors to recent developments. A new 
perspective gained is the surprising possibility that an important part of gauge-theory 
dynamics may be encoded into spacetime physics 
in higher dimensions. On the other hand, 
from the viewpoint of string theory, gravitational physics in the 
bulk is equivalently described by lower dimensional 
systems associated with various  branes. It is important 
to understand this correspondence, including its origin, applications, 
and extensions, as deeply as possible from complementary  standpoints  of field theory and of string theory. 

However, it has been very difficult to check this correspondence 
when it involves really nontrivial dynamical aspects, since we do 
not yet have any systematic way of relating 
both sides from first principles and therefore  
we would be required to solve the dynamics to the 
extent that is necessary to check the correspondence 
for final physical outcomes.  For instance, we do not have 
appropriate (or {\it intrinsic}) understanding on how the gauge 
principles of both sides, namely, 
general coordinate invariance in the bulk 
and the gauge invariance in internal space 
on the boundary, are related to each other.  Because of 
this difficulty, we could gather some evidence for 
the correspondence only in special circumstances where 
(conformal) supersymmetry protects simple lowest order 
results against various complicated corrections. 

In this situation, it is useful to study appropriate limiting cases 
where we can justifiably use some approximation 
schemes in deriving physical quantities on both sides 
and compare them directly. The large-$J$ limit \cite{bmn} has been one 
of such limits by which some of most interesting results 
have been obtained in recent few years.  It means that 
we consider physical states with very large angular momentum ($J$) 
along some direction in the bulk, and,  
correspondingly, gauge-invariant
 operators with large R-symmetry charges  
in the boundary theory. In the bulk,  we can then have 
certain semi-classical treatments for string states,  
while on the boundary, as it turns out,  we can apply perturbation 
theory with respect to a rescaled 
coupling constant $\tilde{\lambda} \sim g^2_{{\rm YM}}N/J^2$ for gauge theories. 

Even in such limiting cases, however, the 
large majority of investigations \cite{beisert} in this area have  
been concentrated to comparison 
of only the spectrum of conformal dimensions for some special 
sets of BPS or non-BPS operators. In spite of 
its importance, discussions on correlation functions have 
been very scarce.   In this exposition, we first 
review main lines of our works \cite{dsy}\cite{yone}\cite{asy}\cite{dy}\cite{dy2}, in which we have been trying to put 
 the correspondence in the large-$J$ 
limit to the spirit of the original proposal in \cite{gkp}\cite{witten}  of direct holographic 
relation for correlators.  Clarification of relevant issues from this viewpoint is indeed useful for better understanding of 
the meaning and the role of the large-$J$ limit. 
For example, we have been able to extend the correspondence 
beyond the spectrum of conformal dimensions by proposing a 
concrete relation\cite{dy} that expresses the OPE coefficients 
of non-BPS operators in terms of 3-point interaction vertices 
of string-field theory in the bulk. In the present article, we also give a new 
solution of infinitely extended string-world sheet which should be useful for studying the deformation 
of Wilson-loop operators from the viewpoint of bulk string theory. 
The author hopes that the present compact 
account of our woks and various comments on 
related issues serve the reader as a guide 
to our original works and as a stimulus for 
further investigations. 

\section{How to reconcile PP-wave limit with holography?}
\subsection{Standard argument of the PP-wave limit}
Let us start from recalling the standard argument\cite{bmn} of 
the PP-wave limit and the BMN limit associated with it. 
In the AdS$_5\times$S$^5$ metric using the global coordinate 
\EQ
ds^2=R^2\Big[
-\cosh^2\rho (dt_g)^2 +(d\rho)^2 +\sinh^2\rho d\Omega_3^2
+\cos^2\theta (d\psi)^2 +(d\theta)^2 +\sin^2\theta d\tilde{\Omega}_3^2
\Big], 
\label{adsglobalmetric}
\EN
we first choose a null geodesic
\EQ
t_g=\psi, \quad \rho=\theta=0
\label{nullgeodesic}
\EN
which traverses a large circle of $S^5$  at 
the center of the AdS$_5$ spacetime. 
If we are interested only in a vicinity of the geodesic, 
we can enlarge the region around the trajectory 
by redefining the coordinates as 
$r\equiv R\rho, y\equiv R\theta$, 
$x^+=(t_g+\psi)/2, x^-=R^2(t_g-\psi)/2$ and taking 
the limit of large $R$. The metric then reduces to
\EQ
ds^2=-4dx^+dx^- -x_i^2 (dx^+)^2 +
(dx_i)^2
\label{ppwavemetric}
\EN
where the index $i$ runs from 1 to 8, of 
which the first 4 ($i=1, \ldots, 4$) 
correspond to the part
 $(dy)^2 +y^2 d\tilde{\Omega}_3^2$  of the reduced metric and the 
last 4 ($i=5, \ldots, 8$) 
to $(dr)^2 + r^2d\Omega_3^2$. If we choose the 
light-cone coordinate $x^+$ as the world-sheet time $\tau$, 
the metric is
 quadratic with respect to 
these transverse coordinates. 
We can then treat the quantum theory of strings 
with this background in terms of a free field theory on the 
world sheet as in flat spacetime. 

A natural question is when this approximation 
is justified. Let us first study  the center-of-mass degrees of 
freedom of strings moving along the null geodesic. The angular momentum along $\psi$ is 
$
J=\alpha R^2 {d\psi \over d\tau}
$
with $\alpha$ being the string-length ($0\le \sigma \le  2\pi \alpha$) 
parameter in the sense of world-sheet in the 
light-cone gauge. The null geodesic can be 
regarded as a point-like classical solution of string 
equation of motion:
$
t_g=\psi={J\over \alpha R^2}\tau \equiv {\mu\over \alpha'}\tau 
$
or equivalently, $x^+={\mu\over \alpha'}\tau, x^-=x_i=0$  with 
$\mu\alpha=\alpha'J/R^2$.  We can 
consider the limit of large $R^2=\sqrt{g_{{\rm YM}}^2N}\alpha' 
(\gg \alpha')$ with $\mu\alpha$ being kept fixed. 
The limit $R^2/\alpha' \gg 1$ allows us to regard the 
quadratic form of  the metric to be a semi-classical approximation 
to full string theory on the AdS background. 
Note that this limit can be taken with a finitely fixed $\alpha'$, 
since we have an independent parameter $N$ which 
controls the higher-genus effect for a large but fixed AdS radius. 
We can consider 
stringy excitations, just as we do in flat background. 

The bosonic part of the string effective action for the transverse part
 is then 
\EQ
S={1\over 4\pi \alpha'}\int d\tau \int_0^{2\pi\alpha}d\sigma 
\Big[
\Big({\partial x_i\over \partial \tau}\Big)^2 -
\Big({\partial x_i\over \partial \sigma}\Big)^2 -\mu^2 x_i^2
\Big] .
\label{bosonaction}
\EN
As discussed in \cite{met}
\cite{mettseyt}, 
the full light-cone gauge action in the Green-Schwarz 
formalism is supersymmetric 
with 32 generators, as it should be since the 
above limit does not violate supersymmetry of the 
original AdS$\times S^5$ background, owing to the presence of 
the 5-form field strength. 
The latter is 
responsible for a mass term, the partner of the 
bosonic mass term in (\ref{bosonaction}), for the world-sheet 
fermionic fields and reduces in the same limit 
as above to
\EQ
F_{+1234}=F_{+5678}=2\mu .
\EN  
Among the 32 susy generators, a half of  them 
correspond, in the language of boundary theory, 
 to the ordinary maximal global supersymmetry of ${\cal N}=4$ 
super Yang-Mills theory, and the other half correspond to the 
global supertranslation of the massless fermion fields. 

The energy of a single free string
in this background takes the form
\EQ
P^-={1\over \alpha}\sum_{n=-\infty}^{\infty}\omega_n
(a_{ni}^{\dagger}a_{ni}+b_{ni}^{\dagger}b_{ni}), \quad 
\omega_n=\sqrt{n^2 +(\mu \alpha)^2}
\EN
with $n$ being the mode number with respect to the 
spatial world-sheet momentum and $a (b)$-operators come 
from bosonic (fermionic) world-sheet fields. 
As in the flat spacetime, we have to impose the 
level-matching condition, 
\EQ
\sum_{n=-\infty}^{\infty}n(a^{\dagger}_{ni}a_{ni}+b^{\dagger}_{ni}b_{ni})
|\Psi\rangle =0 . 
\EN
In particular, states with only zero mode excitations constructed 
from $a^{\dagger}_{0i}, b_{0i}^{\dagger}$  constitute the 
supergravity states. Namely, the tower of states generated by 
acting fermion operators to each given bosonic 
state that is constructed only by bosonic oscillators 
gives a ${\bf 2^8=128+128}$ multiplet of Kaluza-Klein 
supergravity fields, 
which are 1/2-BPS states. 
The spectrum of these  states  
\[
\Delta ={P^-\over \mu}+J=N+J, \quad 
N=a^{\dagger}_{0i}a_{0i}+b^{\dagger}_{0i}b_{0i}
\]
precisely coincides with that of conformal dimensions 
of the corresponding half-BPS operators of Yang-Mills 
theory.  The ground state $N=0$ corresponds to 
the operator $Tr(Z(x)^J)$ with $Z\equiv (\phi_5+i\phi_6)/\sqrt{2}$ 
representing the basic unit of the U($1$) R-charge 
corresponding to the large circle of the geodesic in $S^5$.  
A `scalar' excitation of $a^{\dagger}_{0i}, \, i=1, \ldots, 4$ 
corresponds for the 1/2-BPS operators to the insertion 
of an  `impurity' field $\phi_i(x)$ in a completely symmetric way. 
Similarly,   a `vector' excitation of $a^{\dagger}_{0i}, \, 
i=5, \ldots, 8$ corresponds to the insertion of a derivation
 with respect to one of 
4 base spacetime coordinates $x_{i-4}$. 

In the case of nonzero-mode operators, the world-sheet 
momentum $n$ is interpreted, in the language 
of boundary theory,  as the discrete momentum 
associated with the  insertions of scalar fields or of base spacetime derivatives 
along the `string` of local operator products of approximate length $J$ 
in the large $J$ limit. 

A crucial observation in \cite{bmn} is that 
the spectrum of $P^-/\mu$ for higher excited states with nonzero 
mode $n$ gives the anomalous dimensions of non-BPS operators 
which are obtained by general {\it non}-symmetric 
and a finite number of insertions of impurity  
scalar fields and of base-spacetime derivatives. 
This has been explicitly confirmed to all orders within the 
planar approximation, at least for
 the case of 
scalar excitations, in 
perturbative expansion of the energy
\[
\omega_n=\sqrt{n^2 +(\mu\alpha)^2}=\mu\alpha +
{n^2 \over 2\mu\alpha} + \cdots
\label{confdim}
\]
with respect to $1/(\mu\alpha)^2=R^4/J^2=g_{{\rm YM}}^2N/J^2$. 
Thus we can explicitly see the effect of higher stringy modes 
in the holographic relation between bulk string theory and 
gauge theory on its boundary 
by considering the large $J$ limit appropriately. 

The correspondence has  been extended further to more 
general operators\cite{minahan} in which the number of $Z$ and 
impurities are of the same order of  magnitude. 
The corresponding strings 
in this generalization are interpreted\cite{kurc} as `spinning strings' that are 
characterized by two or more independent angular momenta 
in the bulk, while on the gauge-theory side the 
spectrum of anomalous dimensions is obtained by mapping 
the mixing matrix of anomalous dimensions to a Hamiltonian 
of integrable spin-chains in statistical mechanics: 
different spin states 
in the latter are interpreted as 
different components within the 
strings of local products of scalar fields. 

\subsection{Puzzles and resolution}
Let us now discuss this remarkable correspondence 
from the viewpoint of holographic principle of AdS/CFT 
correspondence. In the standard  
interpretation, the boundary which constitutes a holographic 
screen is located at the conformal boundary $\rho \sim 
\infty$ in terms of the  AdS metric (\ref{adsglobalmetric}). 
Then the (3+1)-dimensional coordinates $(t_g, \Omega_3)$ on 
the  boundary are identified with the R$\times S^3$ foliation 
(or `radial quantization') 
of the base spacetime of super Yang-Mills theory.   
In this sense, energies in the bulk are 
related to conformal dimensions in the boundary 
theory, since evolution operator in the latter can be 
identified with the dilatation operator in the Euclidean 
formulation. 

Comparing 
this interpretation with the above ansatz for the 
correspondence between PP-wave string theory and super 
Yang-Mills theory, we notice apparent contradictions.\cite{dsy}
The BMN ansatz assumes that the 
four ($i=5, \ldots, 8$) of transverse directions which 
must be by definition orthogonal to the time direction are identified with 
the directions of base spacetime. According  to the above idea of 
holography, however, the time direction of bulk and boundary is 
one and the same global time $t_g$. 
Also, while the BMN ansatz requires an Euclidean metric 
of base spacetime for the boundary, the PP-wave limit as 
formulated above clearly 
assumes a Minkowski metric.  As a matter of fact, 
one might forget about the problematical four transverse 
directions by focusing attention only to the scalar impurities. 
But then we would 
lose our way to a systematic understanding 
of correlation functions from the viewpoint  of holography, 
as we shall discuss below. 

A possible attitude 
facing this perplexing situation could be to try to find a different 
interpretation of holography 
and study the boundary\cite{bn} for the PP-wave background {\it per se}, 
independently of its ancestor, AdS/CFT. 
We do not take this viewpoint, since it seems more natural 
to seek for a resolution by trying a reconciliation between 
the coincidence of anomalous conformal dimensions 
and the original interpretation of holography 
connecting AdS bulk and its boundary, 
since after all the BMN operators are nothing other 
than the operators of original Yang-Mills theory. 
 In particular, it 
seems very desirable to keep the direct relation, embodied as the famous GKPW formula, 
between the correlation functions of 
boundary theory and the partition function of 
bulk theory. 

From our standpoint, a crucial problem with 
respect to the holographic principle is that 
the null geodesic (\ref{nullgeodesic}), sitting at the center of the 
AdS, 
 does never reach the 
AdS boundary $\rho=\infty$. Thus one naively inclines to 
conclude that 
it would not be feasible to establish any direct link between the 
PP-wave geometry and the original AdS/CFT correspondence. 
That is not correct. 

To get a useful perspective, it is instructive to consider this 
trajectory using the 
Poincar\'{e} metric $ds_{ads}^2=R^2((dz)^2+(d\vec{x})^2)/z^2$ 
for AdS$_5$ with 
$(d\vec{x})^2$ being the metric of $R^{3,1}$ as the 
base spacetime of the boundary, in terms of which 
the geodesic (\ref{nullgeodesic}) is given as ($t$ being 
the time coordinate of $R^{3,1}$) 
\EQ
z=1/\cos t_g, \quad t=\tan t_g, \quad \psi=t_g. 
\EN
Here for brevity, we dropped the physical parameters $J$ and $E$ 
from the equations.  This is always possible by a suitable rescaling of 
the coordinates. 
The conformal boundary corresponds to $z=0$, and 
$z=\infty$ is the horizon of the Poincar\'{e} patch. 
The periodicity in $t_g$ reflects the fact that the AdS$_5$ 
background can be regarded as a universal covering of a hyperboloid 
embedded in $R^{5,1}$. 
In this picture,  the global time coordinate $t_g$ 
plays  the role of the affine parameter along the trajectory.  
The trajectory does not reach the boundary because 
of the restriction $z\ge 1$ arising from the null condition 
(Virasoro condition) 
$(\dot{z}^2-\dot{t}^2)/z^2+\dot{\psi}^2=0$ with 
$\dot{t}/z^2=1$ and $\dot{\psi}=1$. Thus we 
have 
effectively the one-dimensional constraint $\dot{z}^2=z^2(z^2-1)$, 
which exhibits the existence of a 
centrifugal wall at $z=1$ separating the central region $z\rightarrow \infty$ and the 
conformal boundary $z=0$. 

Now, if we make a Wick rotation 
$t_g\rightarrow -i\tau, \, t\rightarrow -i x_4$, 
for the global time and the boundary time simultaneously,  
the solution turns into 
\EQ
z=1/\cosh \tau, \quad x_4=\tanh \tau
\label{tunnel}
\EN
since the null condition becomes $\dot{z}^2=z^2(1-z^2)$. 
Obviously, this  corresponds to considering a 
`tunneling' trajectory on the opposite side $z \le 1$ of the 
wall, and correspondingly the trajectory 
now starts at $\tau=-\infty$ from the boundary  and returns 
at $\tau=+\infty$ to the boundary, traversing a finite distance 
$\Delta x_4=2$ in the Euclideanized time direction.  
Thus we are led to treat the Euclideanized AdS (EAdS) background. 
The Wick rotation requires us a further rotation 
$\psi \rightarrow -i\psi$ 
along a large circle of $S^5$, which is understandable 
in the large $J$ limit. 

The fact that the bulk-boundary correspondence 
is indeed a tunneling phenomenon is easily seen  
from the behavior of bulk fields near the boundary: 
A massless scalar field in the bulk, for 
instance, behaves as $\phi(z, x)
\rightarrow z^{-J}\phi(x)$ or $z^{J-4}\phi(x)
$ near the boundary. According to \cite{witten}, 
the former {\it non-normalizable} solution gives 
correlation functions on the boundary with conformal 
dimension $\Delta=J+4$. The exponential behavior 
$\exp[\pm J \tau]$ for large $|\tau|$  is indicative of tunneling. 
When we do not consider the PP-wave limit, whether we use 
Minkowskian or Euclidean signature is more or less 
a matter of convention and is not a big problem 
for usual purposes. However, after the limit being taken, 
the topological character of the limiting 
spacetime in relation to the original AdS background 
critically depends on these choices. If one only considers the 
regions near the turning points $\tau\sim 0$ or $t_g\sim 0$, 
both of these trajectories pass through essentially the 
same region of the AdS spacetime. In this latter sense, 
both pictures may be connected. In particular, the free 
spectrum of strings in the PP-wave limit should be 
derivable equally from either viewpoint. But the tunneling 
picture is crucial for discussing correlation 
functions from the viewpoint of GKPW relation. 

Near the boundary, a translation $\tau 
\rightarrow \tau +c$  
with respect to the affine time is equivalent to the scaling 
$z\rightarrow \e^{\pm c}z$ of radial coordinate 
of the EAdS  background. Because of the conformal isometry 
of the EAdS, this scaling is equivalent to a dilatation 
in $R^4$ on the boundary. This naturally explains 
the fact that 
the spectrum of the Hamiltonian for 
evolution with respect to the affine time coincides with 
that of the dilatation operator on the Yang-Mills side. 
This also gives a resolution for the question why we can identify the 
half ($i=5, \ldots, 8$) of transverse directions with the 
base spacetime directions at the boundary, since 
now the direction of the tunneling trajectory is 
manifestly orthogonal to the boundary. 
Thus we have a very simple and natural resolution of the 
puzzles of the BMN correspondence and its 
holographic interpretation. It is straightforward to 
check that we obtain essentially the same 
metric  and effective action for 
string propagation by blowing up the spacetime along the 
tunneling geodesic as (\ref{ppwavemetric}), with 
 an appropriate rearrangement of variables. 
Note that the light-cone time in this  formulation 
flows along the tunneling trajectory. 
We refer the reader to \cite{dsy}\cite{dy} for details of such 
calculations. 

\section{Large $J$ limit of the GKPW relation}
The direct connection of 
the tunneling null geodesic to the EAdS boundary enables us 
to study the large $J$ limit using the GKPW relation
\EQ
Z[\phi]_{{\rm bulk}} \sim \left\langle \exp \Big[\int d^4 x \sum_i\phi_i(\vec{x})O_i(\vec{x})\Big]
\right\rangle_{{\rm boundary}} 
\label{gkpw}
\EN
and thereby to examine the behavior of the holographic 
relation of correlators explicitly.  Remember that this type of 
correspondence has further been 
generalized to (generalized) Wilson loop operators in \cite{rey}
\cite{maldacena2}  and has led to many intriguing results. 
It is a matter of course that when we consider the limit 
of small Wilson loops the prescription of computing 
correlators of Wilson loops from the bulk viewpoint 
must reduce to (\ref{gkpw}).

This conjectural relation has been essentially confirmed at least 
in the case of two-and three-point functions of chiral primary   operators by explicit computations on both sides. 
On the bulk side, the correlators are obtained 
diagrammatically by the so-called Witten diagram. 
For instance, in the leading planar approximation, 
a 3-point  function of scalar operator 
takes the form
\[
\int {d^4\vec{x} dz 
\over z^5} \, 
K_{\Delta_1}(z, \vec{x};\vec{x}_1)
K_{\Delta_2}(z, \vec{x};\vec{x}_2)
K_{\Delta_3}(z, \vec{x};\vec{x}_3) 
\]
where 
\EQ
K_{\Delta}(z, \vec{x};\vec{y})
={\Gamma(\Delta)\over 
\pi^2 \Gamma(\Delta-2)}
\Big(
{z \over z^2+ (\vec{x}-\vec{y})^2}
\Big)^{\Delta}
\label{bbpropagator}
\EN
is the bulk-boundary propagator of scalar field, 
connecting a bulk position $(z, \vec{x})$ to 
a boundary point $(0, \vec{y})$ and satisfying 
\EQ
\lim_{z\rightarrow 0} z^{\Delta -4}K_{\Delta}(z, \vec{x};\vec{y})
=\delta(\vec{x}-\vec{y}). 
\label{bcond}
\EN 

Since, in the large $J$ limit, the conformal dimension is also of order 
$J$, we can apply saddle-point method\cite{dy} 
in evaluating the integral in the above expression. Let us first 
consider the simpler case of a 2-point function. 
\EQ
G_2(\vec{x}_1, \vec{x}_2) \equiv \int {d^4\vec{x} dz 
\over z^5} \, 
K_{\Delta}(z, \vec{x};\vec{x}_1)
K_{\Delta}(z, \vec{x};\vec{x}_2)z^{\epsilon}, 
\label{defG2}
\EN
where the factor $z^\epsilon$ is introduced as a regularization. 
The saddle-point equation is then 
\EQ
{\partial \over \partial z}
\Big(
\ln [{z \over z^2 + (\vec{x}-\vec{x}_1)^2 }]
+
\ln [{z \over z^2 + (\vec{x}-\vec{x}_2)^2 }]
\Big) =0, 
\EN
\EQ
{\partial \over \partial \vec{x}}
\Big(
\ln [{z \over z^2 + (\vec{x}-\vec{x}_1)^2 }]
+
\ln [{z \over z^2 + (\vec{x}-\vec{x}_2)^2 }]
\Big) =0.
\EN
It is easy to see that the general solution 
takes the form
\EQ
z_{saddle}={|\vec{x}_1-\vec{x}_2|\over 2\cosh\tau}
,\quad \vec{x}_{saddle}={1\over 2}(\vec{x}_1+\vec{x}_2)
-{1\over 2}(\vec{x}_1-\vec{x}_2) \tanh \tau
\label{saddle}
\EN
with $\tau$ being an undermined integration constant.  
Evidently, this is precisely the tunneling 
trajectory (\ref{tunnel}), after making suitable 
scaling, shift and rotation of coordinates, provided that we identify the 
`collective' coordinate $\tau$ with the affine time. 
This implies that in the large $J (\sim \Delta)$ limit 
the bulk spacetime can be effectively replaced 
by the vicinity around a tunneling trajectory  
which connects two points $\vec{x}_1, \vec{x}_2$ of 
AdS boundary where the 
local operators are inserted, as it should be from the 
discussion of the previous section. 

The  integration over the collective coordinate and 
over Gaussian fluctuations around the 
trajectory can be performed in a standard way. 
The measure factor is transformed as 
\EQ
{dz d^4\vec{x} \over z^5} \Rightarrow d\tau
d\tilde{z}d^3\vec{\tilde{x}}_{\perp}J(\tau)
\label{measure}
\EN
with $\tilde{z}$ and $\vec{\tilde{x}}$ being the 
linearized fluctuation in the directions 
orthogonal (${dz_{saddle}\over d\tau}\delta\tilde{z}
+{d\vec{x}_{saddle}\over d\tau}\cdot \delta\vec{\tilde{x}}=0$) to trajectory and  
\EQ
J(\tau)=\sqrt{\Big( {4 \cosh^2\tau\over |\vec{x}_1-\vec{x}_2|^2}
\Big)^4 \cosh^2 \tau} . 
\EN
It turns out that the Jacobian factor $J(\tau)$  is precisely cancelled 
by the determinant coming from the integral over fluctuations, and 
the final result takes the form
\EQ
 G_2(\vec{x}_1, \vec{x}_2)\Rightarrow 
{\Delta^2\over \pi^2}|\vec{x}_1-\vec{x}_2|^{-2(\Delta-\epsilon)}
\int_{-\infty}^{+\infty} d\tau \, (2\cosh \tau)^{-\epsilon} .
\EN
By making a suitable renormalization  in 
the limit $\epsilon \rightarrow 0+$, this gives the correct 
two-point function. 

It seems clear that in order to utilize the same {\it single} tunneling 
trajectory for two-point functions for higher-point functions too, 
we have to arrange the points of operator insertions at the 
boundary $z\sim 0$ into two small bunches. Each 
insertion must be located close  to either $\vec{x}_1$ or $\vec{x}_2$.  
In the case of a 3-point function, we can choose $\vec{x}_2$, 
renamed as $\vec{x}_c$, to be the 
 center-of-mass point of two operator insertions 
(with R-charges $J_2$ and $J_3$ satisfying $J_1=J_2+J_3$) 
whose distance is $2\delta$. For sufficiently small $\delta$, 
the same single trajectory as above can be a 
good approximation to the saddle point solution,  
until we approach too close 
the point $\vec{x}_c$. It turns out that, with respect to the 
affine time parameter $\tau$, 
 the range of validity 
of this approximation is controlled by the condition
\[
\e^{-2\tau} > {J_2J_3\over J_1} {(2\delta)^2\over 
|\vec{x}_1-\vec{x}_c|^2}, 
\]
and the final integral form of 3-point function after 
integration over the fluctuations is 
\[
{\pi^2 \over J_1^2}
|\vec{x}_1 -\vec{x}_{c}|^{-(\Delta_1 + \Delta_2 
+\Delta_3)}
\int_{-\infty}^{+\infty} d\tau \, 
\e^{-(\Delta_1-\Delta_2-\Delta_3)\tau}
\exp[-{J_2J_3\over  J_1}{(2\delta)^2\over 
|\vec{x}_1-\vec{x}_c|^2}\e^{2\tau}]
\]
\EQ
={\pi^2 \over J_1^2} \, |\vec{x}_1-\vec{x}_c|^{-2\Delta_1}
(2\delta)^{-(\Delta_2+\Delta_3-\Delta_1)}
\Big({J_2J_3 \over J_1}\Big)^{-(\Delta_2+\Delta_3-\Delta_1)/2}
{\Gamma({\Delta_2+\Delta_3-\Delta_1\over 2}+1)
\over -\Delta_1+\Delta_2+\Delta_3}  . 
\label{3psaddle}
\EN
Note that in our small $\delta$ limit this is consistent with 
the general form of 3-point functions of operators with 
definite conformal dimensions with OPE coefficient $C_{123}$, 
\EQ
{C_{123} \over 
|\vec{x}_1-\vec{x}_2|^{2\alpha_3}
|\vec{x}_2-\vec{x}_3|^{2\alpha_1}
|\vec{x}_3-\vec{x}_1|^{2\alpha_2}}
\sim {C_{123}\over |\vec{x}_1-\vec{x}_c|^{2\Delta_1}
|2\vec{\delta}|^{2\alpha_1}}
\EN
where $\alpha_1=(\Delta_2+\Delta_3 -\Delta_1)/2$ etc. 

It is instructive to rewrite (\ref{3psaddle})  
in the form 
\[
{\epsilon^{\Delta_1-\Delta_2-\Delta_3}
\over -\Delta_1+\Delta_2+\Delta_3}
\times 
\Big({J_2J_3\over J_1}\Big)^{{\Delta_1-\Delta_2-\Delta_3\over 2}}\Gamma({-\Delta_1+\Delta_2+\Delta_3\over 2}+1), 
\quad \epsilon=2\delta.
\]
The first factor has been predicted on the basis of 
the tunneling picture in our first 
paper\cite{dsy} in which, owing to an ambiguity 
of regularization,  we could not have derived the 
second factor originating from the exponential 
cutoff factor in the integral form (\ref{3psaddle}). 
The cutoff factor leads to the fact that the energies 
(conformal dimensions) are not conserved 
even though we are considering `S-matrix' 
along the tunneling trajectory. 
By contrast, if we have naively considered S-matrix in the 
PP-wave background in the standard interpretation, 
energies must be 
conserved, and hence we would not be able to 
relate scattering matrix elements in the Minkowski signature to 
the correlation 
functions of gauge theory in any meaningful manner.

Thus we have now established a concrete 
way of dealing with correlators in the PP-wave 
background by reconciling the large $J$ limit with  
the original formulation of AdS/CFT holography. 
We would like to invite the interested readers to 
our original papers\cite{dy}\cite{dy2} which contain all the details and 
necessary extensions 
of the above arguments. 

For scalar chiral primary operators, the precise relation 
between the OPE coefficient and the three-point 
interaction vertex of effective field theory in the bulk has been 
known from the work\cite{lmrs}. The bulk effective action is 
\EQ
S_{bulk}={4N^2\over (2\pi)^5}\int d^5 x \sqrt{g}
\Big[\sum_{I} {1\over 2}(\nabla\phi_I)^2 
+ {1\over 2}k(k-4)\phi_I^2 
-{1\over 3}\sum_{I_1,I_2, I_3}
{{\cal G}_{I_1I_2I_3}\over 
\sqrt{A_{I_1}A_{I_2}A_{I_3}}}
\phi_{I_1}\phi_{I_2}\phi_{I_3}\Big]
\label{bulkaction}
\EN
where we used Euclidean metric and assumed 
a particular normalization for the scalar fields:
\EQ
A_I=2^{6-k}\pi^3 {k(k-1)\over (k+1)^2} , 
\EN
where $k=\Delta=J+\tilde{k}$ is the conformal 
dimension with $\tilde{k}$ being the number of 
impurities exciting in the 4 directions $i=1, \ldots, 4$. 
In our  large $J$ limit, the known relation between the 3-point 
vertex to the OPE coefficient reduces to   
($\tilde{\Sigma}=\tilde{k}_1+\tilde{k}_2+\tilde{k}_3$) 
\EQ
C^{I_1I_2I_3}={1\over N}{2^{J_1 +{\tilde{\Sigma}\over 2} -9}\over \pi^3}
{\sqrt{J_1J_2J_3}\over J_1^2} \Big({J_1 \over J_2 J_3}\Big)^{\alpha_1}\, 
{\alpha_1! \over \alpha_1}\,  {\cal G}_{I_1I_2 I_3} .
\label{couplingcft2}
\EN
It is easy to check that, 
apart from the normalization factor which is independent 
of $\alpha_1=(\Delta_2+ \Delta_3 -\Delta_1)/2 
=(\tilde{k}_2+\tilde{k}_3-\tilde{k}_1)/2 \equiv \tilde{\alpha}_1, 
$
the $J$ dependence of this relation coincides with the result of our saddle-point computation (\ref{3psaddle}). In particular, 
the prefactor ${\alpha_1!\over \alpha_1}\propto 
\Gamma(1+(\Delta_2+\Delta_3-\Delta_1)/2)/(\Delta_2+\Delta_3-
\Delta_1)$ is responsible for the fact that the 
CFT coefficient $C^{I_1I_2I_3}$ can be {\it finite} 
even when the 3-point vertex ${\cal G}_{I_1I_2I_3}$ 
{\it vanishes} in the extremal case $\alpha_1=0$. 
By this check, we can now fix the absolute normalization 
of 3-point correlators in relating 3-point correlators on the 
boundary to 3-point vertex of the effective field theory in the bulk.

\section{Holographic relation and string field theory in the BMN limit}
The foregoing discussion is a basis for our proposal\cite{dy}  
of general holographic relation between bulk string field 
theory and the OPE coefficients of BMN operators. 
First we derive the effective theory 
around the tunneling geodesic in  supergravity 
approximation, by making a redefinition 
of scalar field in $S_{bulk}$ as 
\EQ
\phi(\tau, \vec{y})
={(2\pi)^{5/2}\over 2N}{1\over \sqrt{2J}}\e^{-J\tau}
\sum_n\phi_n^{(J)}(\vec{y})
\exp[-{1\over 2}J\vec{y}^{\,2}]
\psi_n(\tau)
\label{reduction1}, 
\EN
\EQ
\overline{\phi}(\tau,\vec{y})
={(2\pi)^{5/2}\over 2N}{1\over \sqrt{2J}}\e^{+J\tau} 
\sum_n\phi_n^{(J)}(\vec{y})
\exp[-{1\over 2}J\vec{y}^{\,2}]
\overline{\psi}_n(\tau)
\label{reduction2}, 
\EN
where $\phi_n^{(J)}(\vec{y})$ together with the exponential 
factor $\e^{-J\tau -{1\over 2}J\vec{y}^2}$ are the wave functions for the fluctuations along the trajectory whose Hamiltonian is 
the four-dimensional harmonic oscillator, 
\EQ
h=-{1\over 2}\partial_{y}^2 
+ {1\over 2}J^2 \vec{y}^{\,2}. 
\label{harmonicop}
\EN
Note that the $S^5$ part of wave function  is already factorized 
in these expansions.  
Taking the large $J$ limit, we derive the effective action 
\EQ
\int d\tau\sum_{I}\Big[
{1\over 2}(\overline{\psi}_I\partial_{\tau}\psi_I-
\partial_{\tau}\overline{\psi}_I \psi_I
)+  \overline{\psi}_I {\rm :}h{\rm :}  \psi_I
\Big]
+{1\over 2}\int d\tau \sum_{I_1,I_2,I_3}
\lambda_{\overline{I}_1,
I_2, I_3}(
\overline{\psi}_{I_1}\psi_{I_2}\psi_{I_3} + 
h.c.) , 
\label{totalactionscalar}
\EN
\EQ
\lambda_{\overline{I}_1,I_2,I_3}=
{1\over \pi^3N}
2^{-8+J_1}{2^{\tilde{\Sigma}/2}\over J_1^2}
\sqrt{J_1J_2J_3}\, {\cal G}_{\overline{I}_1,
I_2, I_3}, 
\EN
with $J_1=J_2+J_3$. The zero-point energy is precisely 
cancelled by the contribution coming from $\tau$-differentiation of 
the energy factor $\e^{\pm J\tau}$. It is not difficult to extend this 
action by including the remaining 4 directions corresponding to 
the derivatives of BMN operators with respect to the 4 dimensional 
base spacetime. In the oscillator basis, 
the final form of the effective action turns out to be 
\EQ
S_{eff}=S_2 +S_3, 
\EN
\EQ
S_2=\int d\tau \Big[{1\over 2}\langle \overline{\psi}| \partial_{\tau}|\psi\rangle 
-{1\over 2}(\partial_{\tau}\langle \overline{\psi}|)|\psi\rangle +
\langle \overline{\psi}| h_{sv} |\psi\rangle \Big]
\EN
with 
\EQ
h_{sv}=h_s + h_v, \quad h_s={1\over R}\sum_{i=1}^4a_i^{\dagger}a_i, 
\quad h_v={1\over R}\sum_{j=5}^8 a_j^{\dagger}a_j, 
\label{freehsugra}
\EN
\begin{equation}
S_3=
{1\over 2N}\int d\tau  \,  _{(1)}\langle 
\overline{\psi}|   _{(2)}\langle \psi|  _{(3)}\langle \psi| \sqrt{J_1J_2J_3}
(h^{(2)}_s+h^{(3)}_s -h^{(1)}_s)|v_0\rangle 
+ h.c. \, ,
\label{s3zero}
\end{equation}
\begin{equation}
|v_0\rangle =
\exp \Big[-{1\over 2}\sum_{r,s=1}^3\, \Big(
\sum_{i=1}^4
a^{\dagger}_{i(r)}n_{00}^{rs}
a_{i(s)}^{\dagger} 
+\sum_{j=5}^8
a^{\dagger}_{j(r)}n_{00}^{rs}
a_{j(s)}^{\dagger}\Big)
\Big]|0\rangle
\end{equation}
where $n_{00}^{rs}$ are the supergravity approximation of
 the well-known Neumann 
functions of string-field theory 
restricted to the zero-mode part in the zero-slope 
limit $\alpha'\rightarrow 0$:
\begin{equation}
n_{00}^{rs}=
\delta^{rs}-\sqrt{{J_rJ_s\over J_1^2}} , 
\quad 
n_{00}^{r1 }=n_{00}^{1r}=
-\sqrt{{J_r\over J_1}}
\quad \quad \mbox{for}  
\quad r,s =2, 3 \quad \mbox{and} \quad 
n_{00}^{11}=0. 
\end{equation}
The most characteristic feature of this result is that 
the pre-factor in the 3-point vertex does 
not contain the Hamiltonian $h_v$ of the vector excitations. 
This conclusion is obtained\cite{dy} by a detailed 
examination of descendents of chiral primary operators 
with respect to base spacetime derivatives.  

The emergence of string field theory 
in the zero-slope limit is expected, since we know 
that supergravity approximation is sufficient to reproduce the 
correlator in the case of chiral primary operators. 
To treat non-BPS operators, however, we have to go 
beyond the zero-slope limit. Actually, even for lowest 
supergravity modes, the Neumann functions\cite{pank}  
are subject to $\alpha'$-corrections as
\[
n_{00}^{rs}\rightarrow N^{rs}_{00}=fn_{00}^{rs} 
\quad  \mbox{for}\, \,  \, r=2, 3, \quad 
f=1+4\mu (\alpha')^3{J_1J_2J_3\over 
(\mu R)^3}K 
\]
with $K$ being a complicated function 
expressible in terms of the Neumann coefficient. 
In the limit of large $\mu$ with finite $\alpha'$, $f$ 
behaves as $ {R^2J_1\over 4\pi \alpha' J_2J_3}$.  
Thus, the holographic relation should be 
generalized even for chiral primaries to finite $\alpha'$ such that the 
OPE coefficients for the chiral primary operators are 
not modified but the 3-point vertex are 
replaced by string field theory of finite $\alpha'$ corresponding 
to the above replacement. 

In this way, we are led to propose the following ansatz 
for the holographic relation between OPE coefficient and 
3-point vertex of string field theory in the PP-wave background 
in our Eculidean or tunneling picture:
\begin{equation}
C_{123}=
{\tilde{\lambda}_{123}
\over \Delta_2+\Delta_3-\Delta_1} , 
\label{clambdarelation1}
\end{equation}
\begin{equation}
\tilde{\lambda}_{123}
=\Big(f {J_2J_3\over J_1}\Big)^{-(\Delta_2+\Delta_3-\Delta_1)/2}\Gamma({\Delta_2+\Delta_3-\Delta_1 \over 2}+1)
\lambda_{123}
\label{clambdarelation2}
\end{equation}
where $\lambda_{123}$ 
is the 3-point interaction vertex of string field theory 
whose kinetic part is normalized in the same way as 
the supergravity effective action given above. 
For chiral primary operators in the left hand side 
and hence for states 
with only zero-mode oscillators in the right hand 
side, the additional factor of $f$ 
in this ansatz is precisely cancelled by the modification 
of Neumann functions as above. However, 
for non-BPS states and hence for higher excited oscillators, 
the factor $f$ plays nontrivial roles\cite{dy2} in relating 
bulk and boundary.

It is important here to emphasize that our discussion in fact puts also 
a nontrivial constraint in exploring the correct 3-point interaction vertex 
of string field theory. Namely, for states with only zero-mode 
excitations, 
the 3-point vertex must be equal to 
\begin{equation}
\lambda_{123}
=\,  _{(1)}\langle 1|   _{(2)}\langle 2|  _{(3)}\langle 3| {\sqrt{J_1J_2J_3}\over N}\, 
R(h^{(2)}_s+h^{(3)}_s -h^{(1)}_s)|V_0\rangle , 
\label{clambdarelation3}
\end{equation}
and the corresponding interaction part of the action is 
\begin{equation}
S_3=
{1\over 2}\int d\tau  \,  _{(1)}\langle 
\overline{\psi}|   _{(2)}\langle \psi|  _{(3)}\langle \psi| {\sqrt{J_1J_2J_3}\over N}
(h^{(2)}_s+h^{(3)}_s -h^{(1)}_s)|V_0\rangle 
+ h.c. \, 
\label{s3zero1}
\end{equation}
with $|V_0\rangle $ being the one obtained from $|v_0\rangle$ 
by replacing the supergravity approximation of 
the Neumann functions by the exact ones $N_{00}^{rs}$. 

The interaction vertex must also respect supersymmetry. 
As has already been obvious in our first work, supersymmetry 
is not sufficient to fix the form of 3-point vertex in string field 
theory in the PP-wave limit. Indeed, there have been different 
proposals. Unfortunately, neither of previous proposals alone  
were consistent with the above condition.  It turned out that 
the particular combination of two different types 
of vertex satisfies the above condition. 
With the standard normalization, the correct choice of the 
vertex state for general string states can 
be expressed as
\EQ
|H_3\rangle_h\equiv {1\over 2}(|H_3\rangle_{SV}+|H_3\rangle_D)
\label{h3}
\EN
where first term is the most familiar vertex discussed in 
\cite{SV} and the second term is the 
one advocated in \cite{DPPRT}. 
If we restrict our attention to states with only zero-mode 
oscillators, this form reduces to 
(\ref{s3zero1}), 
\[
|H_3\rangle_h
\rightarrow (h^{(2)}_s+h^{(3)}_s -h^{(1)}_s)|V_0\rangle. 
\]
Thus the 3-point vertex of full `holographic' string field theory 
should be
\EQ
\lambda_{123}=  \, _{(1)}\langle 1|_{(2)}\langle 2|_{(3)}\langle 
3| {\sqrt{J_1J_2J_3}\over N}|H_3\rangle_h .
\label{holo3point}
\EN

The difference between two terms in (\ref{h3}) lies 
in the choice of the so-called prefactor 
that multiplies the familiar overlap function $|V_0\rangle$ 
representing 
the continuity of string field configurations. In the former choice 
$|H\rangle_{SV}$,   
the prefactor is determined by making a straightforward 
generalization of flat space result. In the latter $|H\rangle_{D}$, 
the prefactor is essentially the energy factor 
$H^{(2)}+H^{(3)}-H^{(1)}$ using the free world-sheet 
Hamiltonian $H^{(r)}$ corresponding to three external lines 
$r=1,2,3$. In fact, this particular form has already been 
suggested in our first paper \cite{dsy}.  
Each of these two proposals has been 
associated with a different way of 
relating quantities between bulk and boundary. 
Our proposal is therefore useful to disentangle 
different  ways of relating bulk and boundary, and 
explain the origins of such seemingly {\it ad hoc} prescriptions 
on the basis of the original AdS/CFT correspondence 
as discussed detail in \cite{dy}. 

References \cite{dy}\cite{dy2} presented 
nontrivial explicit checks of our holographic relation 
formulated as the formulae 
(\ref{clambdarelation1}) $\sim$ (\ref{holo3point}) in the 
leading large $\mu$ expansions. 
In particular, it was confirmed that the above holographic 
relation is valid for the case of impurity {\it non}-preserving 
3-point amplitudes at least to our leading expansion, 
as well as for the cases of preserved impurities.  
The structure of the above particular combination of possible 
prefactors of string field theory and also  the additional factor $f$ in the holographic relation play key roles in obtaining these results.  
We emphasize that,
 with respect to this universality of the relation of  OPE coefficients 
of BMN operators and string field theory, there has been 
no other competing proposal.  Previously, there has been no 
discussion on impurity non-preserving processes. 

It should also be mentioned 
that our particular choice of string-field theory vertex may 
not be completely unique, since we have motivated 
it by studying the zero-mode part. Although we have 
many nontrivial checks for the cases with small number of 
insertions of impurities with nonzero modes,  they are not yet sufficient 
for excluding the possibility of modifying string-field theory vertex further. In fact, the work\cite{russo} have suggested a possible modification which 
does not affect all of explicit checks mentioned above but may 
affect more complicated insertions of fermionic impurities. 
In principle, of course, it would be possible to derive 
the string field theory in this particular PP-wave background,  
once we have a firm candidate of string field theory 
on the original AdS$_5\times S^5$ background. However, 
in the situation that the latter program is impracticable at the 
present time, our result in a limiting case may hopefully provide a 
useful data. 

\section{Further  applications and extensions}
In this final section, we summarize some applications or  extensions of our ideas, and then discuss remaining issues. 
\subsection{Case of D$p$-brane background}
It is generally believed that the AdS/CFT correspondence 
should be understood as a special typical case of 
more universal string/gauge correspondence. From the viewpoint 
of string theory, the origin of the correspondence is 
the world-sheet string duality between closed strings and 
open strings. The origin of the world-sheet string duality is 
directly neither 
conformal symmetry, nor (spacetime) supersymmetry, 
that are effective in the AdS/CFT case. Moreover, 
in the case of bosonic string, it is known that Witten's version 
of open string field theory in perturbative 
genus expansion covers the whole moduli space 
of Riemann surfaces including the effects of 
exchange of closed strings. Therefore,  
it is not at all unreasonable to imagine the situation 
where we can generalize the correspondence 
 to 
more general {\it non}symmetric backgrounds and 
their boundaries.  Recently, in fact, there has been 
many `phenomenological' attempts towards construction 
of such holographic duals to realistic QCD and its variants. 

Before going down to 
such a level, it would also be interesting to study the case 
of backgrounds produced by general D$p$-branes that have in 
general smaller supersymmetries and have no 
conformal symmetry in the usual sense. 
For instance, in the case of $p=0$, the gauge theory 
is nothing but the Yang-Mills super quantum mechanics which 
can be interpreted as a discretized light-cone  formulation 
of M-theory or just as the effective low-energy 
theory of D-particles.  In the work\cite{asy}, an extension of tunneling 
picture to such cases has been investigated in order to predict the behaviors of two-point 
functions for corresponding boundary gauge theories. 

Consider the general D$p$-brane background
\EQ
ds^{2}=q_p^{1/2}
\Big[H^{-1/2}(-dt^{2}+d\tilde{x}_{a}^{2})
+H^{1/2}(dr^{2}+r^{2}d\psi^{2}+r^{2}\cos^{2}\psi 
d\Omega_{7-p}^{2})\Big] 
,
\label{Dpmetric}
\EN 
with $H=1/r^{7-p}$ and $q_p \propto
g_{{\rm YM}}^2N$, where $a$ runs from $1$ to $p$. For general $p(\ne  3)$, 
the  horizon $r=0$ 
is a real singular point. Thus it is not meaningful to consider the 
correspondent of the ordinary null geodesic located at 
the `center' of the background. But for $p<5$, we can 
consider tunneling trajectories after making the 
Wick rotation as in the case of $p=3$, which 
satisfy 
\EQ
\dot{r}=\pm H^{-1/2}\sqrt{\ell^{2}r^{-2}-H}
=\pm  r_{0}^{-(5-p)/2}\sqrt{r^{5-p}-r_{0}^{5-p}}.
\label{eq:drdtau}
\EN
For $p<3$, there is a difference from the case $p\ge 3$ in that 
in terms of the natural affine parameter used in the 
above equation, the time interval $2T_b$ connecting boundary 
($r=\infty$) and boundary is finite, 
 \[
T_b=\left({2\over 5-p}\right)^{7-p\over 5-p} 
\int^{z_{0}}_{0} dz 
{z^{-{2\over 5-p}}\over \sqrt{z_{0}^{2}-z^{2}}}
\]
where the integration variable $z$ is related with the original 
radial coordinate $r$ by $z=2r^{-(5-p)/2}/ (5-p)$. But conceptually 
this is not an obstacle for applying our basic ideas. 

The background metric in the PP-wave limit can be 
obtained by the similar methods as in the case of D$3$-branes, 
and the effective string action for fluctuating string 
coordinates turns out to be 
\begin{equation}
S^{(2)}={1\over 4\pi}\int d\tau \int_{0}^{2\pi \alpha}
\hspace{-0.5cm}d\sigma
\Big\{ \dot{\tilde{x}}_{a}{}^{2}+\tilde{x}_{a}'{}^{2}
+m^{2}_{\tilde{x}}(\tau)\tilde{x}_{a}^{2}
+\dot{x}{}^{2}+x'{}^{2}
+m^{2}_{x}(\tau)x^{2}
+\dot{y}_{l}{}^{2}+y_{l}'{}^{2}
+m^{2}_{y}(\tau)y_{l}^{2} \Big\},
\label{eq:Dpfluc}
\end{equation}
where\cite{gimonetal} 
\begin{eqnarray}
m_{\tilde{x}}^{2}&=&m_{x}^{2}=
-{(7-p)\over 16r^{2}}
\left\{(3-p)+(3p-13)\ell^{2}r^{5-p}\right\},\\
m_{y}^{2}&=&
-{(7-p)\over 16r^{2}}
\left\{(3-p)-(p+1)\ell^{2}r^{5-p}\right\}.
\end{eqnarray}
Here the $p$-dimensional vector $\tilde{x}_a$ can roughly be regarded as the fluctuations in the 
spatial directions of D$p$-brane, $x$ in the direction 
of Euclideanized time, and the $(7-p)$-dimensional vector $y_{\ell}$ 
in the directions transverse to D$p$-branes. 
Comparing to the case of D$3$-branes, a complication is 
that the world-sheet mass parameters now depend on the 
affine time through the function $r=r(\tau)$ of the above classical 
solution. 
For $p<3$, the masses asymptotically 
increase indefinitely as we approach the boundary, 
while for $5>p>3$ they vanish 
asymptotically. This leads\cite{asy} to physical consequences that are 
drastically different  from the conformal case,  but are consistent with expected IR properties of non-conformal Yang-Mills theory. 
 
We can apply the same picture as the case $p=3$ to compute 
2-point functions by semiclassical quantization 
of string theory around the tunneling trajectory. 
In \cite{asy}, we have developed a systematic method 
for dealing with the time dependence of mass 
for the purpose of deriving S-matrix along the 
tunneling trajectory. 
The general form of two-point functions 
for operators involving only zero-modes obtained there is 
\EQ
\langle O(x_1)O(x_2)\rangle 
\sim (|x_1-x_2|\Lambda)^{-{4\over 5-p}J
-2N_0^{x} -{4\over 5-p}N_0^y-2C_0(p)}
\label{dpresult}
\EN
($\Lambda$ being a renormalizable regularization 
parameter which is always necessary in deriving 
two-point functions from the bulk viewpoint) 
in the case  of bosonic excitations. 
For $p\ne 3$, there is  no conformal symmetry, 
but the above general form is consistent with 
a generalized conformal symmetry proposed in \cite{jy}. 
 Here $N_0^{x, y}$ 
denote the number of excitations along the 
directions longitudinal ($x$) or transverse ($y$), 
respectively, to the directions 
of D$p$-branes. The constant $C_0$ which is 
related to the zero-point energy on the world sheet is 
calculated in the second ref. in \cite{asy} 
by taking into account the fermionic excitations to be 
$C_0(p) =-(3-p)^2/2(5-p)$, which does not vanish except for 
$p=3$ indicating that the world-sheet supersymmetry is 
in general violated. 
This result is consistent with some previous results  
obtained on the basis of standard linearized analyses 
of supergravity, in particular with \cite{sy}. 

Extension to the case involving higher 
string excitations has also been discussed in 
the first and third references in \cite{asy} together with some interesting implications 
for the IR behaviors of the boundary  gauge 
theories. 

A suggestive observation related to these results is that 
we can extract effective dimensions  $d_{eff}=(14-2p)/(5-p)$ from the 
exponent of (\ref{dpresult}) for transverse 
excitations by interpreting 
 the two-point functions as those of free scalar fields living on 
$d_{eff}$-dimensional effective base spacetime, 
$d_{eff}-2=4/(5-p)$. Of course, for $p=3$, 
we have $d_{eff}=4$ as it should be. Magically 
enough, we have $d_{eff}=3$ or $6$ for 
$p=1$ and $p=4$, respectively. It is tempting to relate this 
phenomena as being a manifestation of 
M2 and M5 branes, respectively,  emerging from the large $N$ limit of 
effective D1 and D4 branes. In fact, in the case $p=1$, 
we can adopt the IIA matrix-string interpretation for D1 branes 
of type IIB theory. Then the Yang-Mills coupling constant 
is reversed, and hence the present weak coupling limit 
must be reinterpreted as the strong coupling M-theory 
limit, that corresponds to a decompactification of 
the radius $R_{11}\propto 1/g^2_{{\rm YM}} 
\rightarrow \infty$ along the 11-th direction. 
In fact, we have discussed in \cite{sy2} how 11-dimensional supermembrane  is  
related to the matrix-string theory. 
Interpretation of $p=4$ case seems more mysterious. 
 
\subsection{Case of spinning-string/spin-chain correspondence}

Let us go back again to the conformal case $p=3$. 
In principle, it is straightforward to extend the analysis 
explained in sections 3 and 4 to the case of spinning strings.\cite{minahan}\cite{kurc} 
Only difference from the BMN limit is the existence of 
two or more independent angular momenta  that 
are sent to infinity. The dominant tunneling trajectory in the 
EAdS background 
describes the center-of-mass motion and therefore can be 
taken to be the same (\ref{tunnel}) or (\ref{saddle}) as in the case of a single angular momentum, 
since its emergence depends only on the limit of large 
conformal dimensions. Therefore, only the treatment 
of the degrees of freedom along the $S^5$ directions 
is affected. The ground state in the case of a single angular 
momentum corresponds to a point-particle solution 
moving along a great circle of the $S^5$. In the case of 
two or more angular momenta, by contrast, we have 
to take into account stringy extension (hence, the naming 
of `spinning strings') even in obtaining classical 
solutions as shown in \cite{frolov}. 

Since almost all of previous discussions of 
spinning strings/spin chain correspondence are focused 
on the comparison of the spectrum between the dilatation operator 
and the string energy, 
it is a useful check to derive two-point correlators directly  
using the same method as the argument in section 4 or 
as in the previous subsection. 
This has recently been done in \cite{tsuji} to which we refer 
the reader. It would also be an interesting exercise to further 
study 3-point functions of spin-chain operators from this viewpoint and 
to investigate how the string field theory of spinning strings 
should look like.  

\subsection{Deformed Wilson loop operators}

Any local operators corresponding to 
those spinning states can be regarded as arising from 
Wilson loops extending along the R-charge directions. 
Thus, computations being discussed here can also be regarded as 
computing the correlators of two small Wilson loops in 
the boundary gauge theory. From this viewpoint, it would also be 
interesting to investigate large Wilson loops and its infinitesimal 
deformations, corresponding to insertions of local gauge-covariant 
operators midst the loops, such as 
\EQ
W_{12, J}=\Tr\Big[{\cal P}
\exp \Big(\oint  d\sigma [A_{\mu}(\vec{x}(\sigma))
\dot{x}^{\mu}
+\phi^i(\vec{x})\sqrt{(\dot{\vec{x}})^2}\theta^i\Big)
 Z(\vec{x}(\sigma_1))^J\overline{Z}(\vec{x}(\sigma_2))^J
\Big]. 
\label{dwilson}
\EN
 Some examples of this type 
have been considered recently in \cite{drukker} for the case 
of circular or straight-line Wilson loops. Reconsidering this 
case from the viewpoint of our tunneling picture seems 
useful for clarifying the holographic interpretation 
in a way consistent with the picture of references.\cite{rey}\cite{maldacena2}\footnote{
In ref.\cite{drukker},  the authors use the `standard' 
picture in the interpretation of the PP-wave limit. We cannot therefore 
use their solution in order to study the 
expectation value of the deformed Wilson loop operator. 
Their solution is subject to no deformation 
near the conformal boundary, since the 
points of deformations are sent to $\pm\infty$ on the  boundary. 
The effect of insertions of local gauge-covariant operators 
is interpreted to correspond to the property that 
solution exhibits rotation around an $S^5$ circle when 
we approach the AdS center.   In fact, one of two patches of the solution 
discussed below is related to their solution by our Wick rotation, 
which drastically changes the topology of the solution as 
emphasized before. 
}

The string world sheet corresponding to  a (doubled) straight-line 
Wilson loop is an infinite plane (actually of double sheet) 
extending from the EAdS boundary to its center 
(and back to the boundary again).  Insertion of a local 
gauge-covariant operator amounts to a 
deformation of the plane at the  vicinity around the 
 point of insertion on the 
boundary. Such local deformations in general propagate into the 
bulk of the world-sheet plane and are expected to produce a similar 
(or perhaps the same) 
trajectory as our tunneling geodesics 
in the limit of large $J$. 

It is indeed not difficult to construct the typical form of  desired 
solution, which represents the particular deformation 
corresponding to (\ref{dwilson}). 
In the conformal gauge, we find 
\EQ
(z, x_4)_{\pm}=
({\sinh \sigma\over \cosh \sigma \cosh \tau \pm 1}, 
{\cosh \sigma \sinh \tau \over \cosh \sigma \cosh \tau \pm 1}), 
\label{adssolution}
\EN
\EQ
(\cos \theta, \psi)=  (\tanh \sigma,  \tau)
\label{s5solution}
\EN
where the angle coordinates $(\theta, \psi)$ are 
those used in (\ref{adsglobalmetric}) with 
the Wick rotation $\psi\rightarrow -i \psi$.  
Here,  the (single) infinite string world sheet is represented by two 
coordinate patches 
corresponding to the sign $\pm$ in (\ref{adssolution}), 
each of which is parametrized in terms of the world-sheet 
coordinates $\tau$ and 
 $\sigma$, running from $-\infty$ and 
$0$, respectively, 
 to $\infty$ and is connected smoothly 
 to each other along a one-dimensional 
 line defined by $\sigma=\infty$. 
The two patches are related by an inversion 
$(z, x_4)\rightarrow ({z\over z^2+x_4^2}, 
{x_4 \over z^2 +x_4^2})$ which is an isometry of the 
EAdS. 
The trajectory along which this  connection is made coincides with our 
tunneling trajectory that is nothing but a unit (half) 
circle  in the $(z, x_4)$ plane; namely, we have
\[
\lim_{\sigma \rightarrow \infty}(z, x_4)_{\pm}
=({1\over \cosh \tau}, \tanh \tau). 
\]
The equation (\ref{s5solution}) together with 
this last property means that, at  $x_4=\pm 1, z=0$ 
corresponding to  the ends (at $\tau=\pm \infty$) of  this 
 trajectory reaching the conformal boundary in EAdS, the string 
rotates around a large circle $\theta=0$ in $S^5$ 
 in the same manner as in the case of 
 subsection 2.2, while, away from these 
two points, $\theta$ approaches the pole ($\theta =\pi/2$) of 
$S^5$, 
corresponding to 
 no rotation.  This is precisely the 
required property corresponding to the above deformation 
occurring at $x_4=\pm 1$ 
of the boundary of the string world sheet which is simultaneously 
at the EAdS boundary. The solution representing the 
plane without deformation is simply given by a single patch 
$(z, x_4)=({\sigma\over \tau^2 + \sigma^2}, {\tau\over \tau^2 + \sigma^2}), (\theta, \psi)=({\pi\over 2},0)$ with the same range of 
world-sheet coordinates. 
It is straightforward to apply similar arguments 
as for the correlators of local operators to obtain the  expectation value 
of the deformed Wilson-loop operator (\ref{dwilson}). 
 More detailed study will be 
reported elsewhere.

\subsection{Bubbling `E'AdS and correlators}
Another interesting path of investigation seems to 
consider the Euclideanized version of the so-called bubbling 
AdS configurations. As is now well known, such 
configurations with 1/2 of the full supersymmetry and with 
SO(4)$\times$SO(4)$\times$R  isometry can be characterized, 
in the Minkowski signature, by droplet configurations\cite{llm}  
defined on a two-dimensional plane embedded in 
the bulk. The case of a single circular droplet, the ground state, 
corresponds to the AdS$_5\times S^5$ background itself. 
The perimeter of this circular droplet coincides with the 
trajectory of the same null geodesic as the one used 
in obtaining the PP-wave limit. This implies that if we 
make the Wick rotation in the way discussed in this 
exposition, the perimeter of circular droplet becomes a hyperbola 
which extends to the boundary.\footnote{
We have already suggested this possibility in \cite{ty2}. 
} Thus  the droplet shows a 
characteristic feature of the old $c=1$ matrix model. 
In particular, the motion of the droplets of general excited states 
propagates to the EAdS boundary. Together with our tunneling 
picture, the amplitudes of this propagation from boundary to 
boundary are expected to give  two-point 
functions of chiral primary operators which correspond to the 
excited states. Namely, the S-matrix of $c=1$ matrix model 
may be related directly to the correlators of chiral primary operators. 
It would be nice to try to formulate this concretely and 
to see to what extent the exact solvability of $c=1$ matrix model 
can guide us the properties of 1/2 BPS operators 
in ${\cal N}=4$ super Yang-Mills theory.  

\subsection{Higher-genus effects}
All our discussions so far have been restricted to the leading 
large $N$ limit, namely the planar limit in the language of 
 gauge theory. There have been 
attempts to derive corrections to conformal dimensions
beyond the planar approximation. On the gauge-theory side, 
the corrections to the formula (\ref{confdim}) is 
computed to be\cite{genus}\cite{genus1}
\EQ
\Delta-J=2(1+{\tilde\lambda} n^2 + \cdots) 
+{J^4\over 4\pi^2N^2}({1\over 12}+{35\over 32\pi^2n^2})(
\tilde{\lambda} + \cdots), \quad 
\tilde{\lambda}={g_{{\rm YM}}^2N\over J^2}.
\label{genusone}
\EN
The second term is the first genus correction. 
Unfortunately, even this leading correction has never been 
derived correctly from the viewpoint of the bulk string theory. 
In fact, a claim has been made 
that this correction could be 
obtained by assuming the particular string vertex 
that retains only the first term $|H_{SV}\rangle$ of 
(\ref{h3}). However, a general consensus at the present time 
is that this was not correct. Aside from some errors, this type of 
calculations is based on an unjustified  basic assumption 
that the sum over the intermediates states in 
the second order calculation is truncated up to excited 
string states with only 2 impurities. 

In a more recent 
computation\cite{pank2} to which we refer the reader for 
more detailed references on previous works, 
 it has been clarified that these assumptions do
 not only 
lead to wrong coefficients, but also to nonperturbative 
terms of the form $\tilde{\lambda}^{1/2}, \tilde{\lambda}^{3/2}, 
\ldots$, which cannot appear in usual perturbative 
expansions on the gauge-theory side. 

Most recently, a calculation using our proposal (\ref{h3}) instead of the 
term $|H_{SV}\rangle$ alone, under the 
same truncation, has been reported in \cite{semenoff}.  According to 
this computation, the situation is  improved with respect 
to the coefficients at the first order in $\tilde{\lambda}$. 
In addition to this,   unwanted nonperturbative 
terms of low orders which are found in \cite{pank2}  are also eliminated by `miraculous' cancellations, 
 occurring by this particular form of the interaction vertex. 
 But the complete 
agreement with the gauge-theory result is still {\it not} attainable 
if one considers $\tilde{\lambda}^2$ terms and highers. 
Therefore, we should definitely take into account other possible 
contributions involving more impurities.

 Another possible source 
of disagreement is that in principle we do not yet fixed 
the next order string-interaction term of the string action. 
The discussion of the string vertex reviewed in section 4 
is effective only at the first order with respect to the 
string coupling. As emphasized in our works \cite{dsy}\cite{dy}, 
supersymmetry is not sufficient to fix the higher order 
corrections to the interaction Hamiltonian. Higher genus 
corrections would crucially depend on the structure of 
the higher order interaction terms. For example, if the interaction 
Hamiltonian were obtained solely by a unitary transformation from  the 
free Hamiltonian, the conformal 
dimensions would not be corrected by higher 
genus effect. The second part $|H_3\rangle_D$ of the 
interaction term in (\ref{h3}) indeed is of this form. 
Similarly, some part of the higher interaction terms 
may also be of this form, as suggested in our first paper\cite{dsy} 
in a different context. 
This might also be somehow related to the truncation 
of higher impurity states, if that is after all destined to be  
justified. 

We can also raise a subtle question as to the dependence on the  
order of various limits involved in this problem. 
For instance, from the viewpoint 
of string theory in the AdS background, it is not at all 
obvious whether we can take the large $J$ limit 
before completing summation over infinite number of 
intermediate states in higher-order calculations, 
since two-body states with a large total angular momentum 
$J$ can consist 
of  one-body states with large angular momentum of 
the same order as $J$  and those with smaller 
angular momentum of order $O(1)$. For the 
latter part, we cannot have any {\it a priori} justification 
for using the large $J$ limit.  In any case,   
the fact that the situation is much improved with 
our proposal 
is a good news, but further painstaking but rewarding 
 efforts, hopefully, are required to clarify this issue.

\section*{Acknowledgements}

The author would like to thank the organizers of the 
symposium for invitation and for patiently waiting 
the present manuscript. Thanks are also due to 
S. Dobashi and A. Miwa for pointing out typos 
in a preliminary version of the manuscript. 
The present work is supported in part by Grant-in-Aid for Scientific Research (No. 13135205 (Priority Areas) and No. 16340067 (B))  from the Ministry of  Education, Science and Culture, and also by Japan-US Bilateral Joint Research Projects  from JSPS.

\end{document}